# Estimating the coefficients of variation of Freundlich parameters with weighted least squares analysis

J.J.T.I. Boesten

Wageningen Environmental Research, PO Box 47, 6700AA Wageningen, the Netherlands; jos.boesten@wur.nl; phone: +31 317 481620

**Abstract**

The Freundlich isotherm has been used widely to describe sorption of solutes to soils for many decades. The Freundlich parameters are often estimated using unweighted least squares (ULS) analysis after log-log transformation. Estimating the accuracy of these parameters (characterized by their coefficient of variation, CV) is an essential element of the use of these parameters. Accurate CVs can be derived with weighted least squares (WLS), but only if proper weights are assigned to the residuals in the fitting procedure. This work presents the derivation of an analytical approximation of these weights which were found to decrease with increasing concentration, increasing Freundlich exponent, and increasing CVs of the initial and equilibrium concentrations. Monte-Carlo simulations for a wide range of Freundlich systems based on known values of the CVs of the initial and equilibrium concentrations confirmed the accuracy of this analytical approximation. Unfortunately, in practice, the CVs of the initial and equilibrium concentrations are unknown a priori. Simulations showed that the accuracy of the estimated CVs of the Freundlich parameters was distinctly lower if the CVs of the initial and equilibrium concentrations were estimated from the isotherm data. However, this accuracy was still considerably better than when using ULS. It is recommended to use this analytical approximation whenever these CVs are relevant for the further use and interpretation of the estimated Freundlich parameters.

# 1 Introduction

The Freundlich equation (Freundlich, 1907) has been used widely to describe sorption of solutes such as phosphate, potassium, heavy metals, hydrocarbons, and pesticides to soils and soil components such as clay minerals, humic acids, oxides etc (e.g. Travis & Etnier, 1981; Che et al., 1992; Buekers et al., 2008; He et al., 2011; Pronk et al., 2013; Toor & Sims, 2015). The Freundlich equation can be derived assuming a system with heterogeneous sorption sites (Sips, 1950). The derivation requires that only a small fraction of the available sorption sites are occupied with sorbate molecules (Van Riemsdijk et al., 1986). Thus, the widespread use of the Freundlich equation in environmental systems likely arises from the fact that most soil-related sorbents systems have heterogeneous sorption sites and that the sorption levels of the sorbates in soils are typically considerably lower than those corresponding with full occupation of the sorption sites.

Thus, adequate assessment of the parameters of the Freundlich isotherm and their uncertainty is an important issue. These parameters are obtained by a fitting procedure using either the log-log transform or the original Freundlich equation. Many publications use this log-log transformation (e.g. Gaillardon et al., 1977, Stougaard et al., 1990; Boesten & van der Pas, 2000; Buekers et al., 2008; Vega et al., 2008; Mouni et al., 2009; Larsbo et al., 2009; He et al., 2011; Lado et al., 2013; Jones & Loeppert, 2013; Kadyampakeni et al., 2014; Toor & Sims, 2015) and therefore, this option was used in this work.

Least squares (LS) analysis commonly assumes that the dependent variable '$y$' is a function of an error-free independent variable '$x$' (Draper & Smith, 1998). However, in sorption studies, both $x$ and $y$ have errors. The standard approach for the log-log transformed Freundlich equation is to assume that the amount sorbed is the dependent variable and that the concentration in the liquid phase is the error-free independent variable (further called 'the standard model'). An alternative approach is to assume that both the amount sorbed and the concentration in the liquid phase have experimental errors (further called 'the errors-x&y model'). Additionally, the weighting of the residues of the fits using either unweighted least squares (ULS, also called 'ordinary' least squares) or weighted least squares (WLS) based on a given error model must be considered. The estimated values of the Freundlich parameters are usually hardly influenced by the selected model and selected weights (Bolster & Tellinghuisen, 2010; Boesten, 2015). The reason for this lack of influence is likely that adequate experimental procedures that prevent large scatter in the data are commonly used, in which case any LS model is likely to perform well (provided that the data conform to a Freundlich isotherm). However, one should consider not only the estimated parameter values but also the assessment of their uncertainty, i.e., their estimated coefficient of variation (CV). To obtain adequate CVs, it is necessary to give the *x-y* pairs (further called 'sorption points') proper weights in a WLS procedure, resulting in the correct $\chi^2$ distribution of the sum of squares (Bolster & Tellinghuisen, 2010). Researchers that fit their data to a Freundlich equation after log-log transformation typically do not report their LS procedure (e.g. He et al, 2011). This lack of reporting likely means that they used ULS. Nevertheless, these CVs are occasionally used when drawing conclusions. Therefore, the accuracy of the estimated CVs of the Freundlich isotherm parameters should be given further consideration.

As discussed by Tellinghuisen & Bolster (2010) and Bolster & Tellinghuisen (2010), use of the errors-x&y model is rather complex. Therefore, they recommend using the standard model in combination with an 'effective variance' approach to estimate the WLS weights. This work presents an analytical approximation for the weights to be used for WLS analysis after log-log transformation using this effective variance approach. Monte Carlo simulations are used to demonstrate that this analytical approximation results in better estimation of CVs of Freundlich parameters than the ULS procedure.

## 2    Model for estimating Freundlich parameters with WLS

The Freundlich equation can be written as:

$$X = K_F \, c_{ref} \left( \frac{c_e}{c_{ref}} \right)^N \tag{1}$$

where $X$ is the mass of sorbate sorbed per mass of sorbent (mg/kg), $K_F$ is the Freundlich sorption coefficient (L/kg), $c_e$ is the mass concentration in the liquid phase of the sorbate after equilibration (mg/L), $c_{ref}$ is a reference concentration in the liquid phase (introduced to prevent the unit of $K_F$ from becoming a function of $N$ and commonly set to 1 mg/L; Boesten, 2007) and $N$ is the Freundlich exponent describing the curvature of the isotherm which is a measure of the heterogeneity of the sorption sites: $N = 1$ (linear isotherm) indicates no heterogeneity and a smaller $N$ indicates more heterogeneity (Kinniburgh et al., 1983). The notation '$N$' was used as this notation is more straightforward than the '1/*n*' notation. This notation is a recurring point of debate (Barrow, 2008) and may lead to confusion (Bowman, 1982), so let us consider the source of the '1/*n*' notation. Freundlich (1907) wrote Equation 1 as $x/m = \beta \, c_e^{1/n}$, referring to five earlier publications (1894-1905) which used this equation (called 'the exponential expression' by Freundlich) to describe adsorption from solutions. These earlier publications wrote the equation as $X^n /c_e$ = constant (Schmidt, 1894) or $\sqrt[n]{c_e} / X$ = constant (Georgievics & Löwy, 1895) which explains the preference for the '1/*n*' notation, which Freundlich (1907) thereafter followed. Because Schmidt and Georgievics & Löwy did not provide any conceptual basis for the use of the 1/*n*-notation, there appears to be no justification for the continued use of 1/*n*.

It is assumed that the sorption is derived from the decrease in the sorbate concentration in the liquid phase, as is assumed in nearly all sorption studies. The *X* values are then calculated as:

$$X = \frac{V(c_i - c_e)}{M} \tag{2}$$

where *V* is the total volume of liquid in the system (L), $c_i$ is the initial mass concentration in the liquid phase of the sorbate (mg/L) and *M* is the mass of sorbent in the system (kg).

As will be shown below, the fraction or percentage decrease in the sorbate concentration in the liquid phase, *δ*, plays a central role in the estimation of the weights. It is defined as:

$$\delta \equiv \frac{c_i - c_e}{c_i} \tag{3}$$

Combining Equations 1, 2 and 3 yields:

$$\delta = \frac{R \, K_F \, C^N}{C + R \, K_F \, C^N} \tag{4}$$

where *R* is the sorbent-liquid ratio (i.e., *M* / *V* in kg/L) and *C* is the dimensionless ratio $c_e / c_{ref}$. Equation 4 indicates that *δ* increases with increasing *R*: if *R* = 0, then *δ* = 0, and if *R* approaches infinity, then *δ* approaches 1, so 100%.

The log-log transformed form of Equation 1 is:

$$\log(X) = \log(K_F \, c_{ref}) + N \, \log\left(\frac{c_e}{c_{ref}}\right) \tag{5}$$

Equation 5 can be interpreted as the following error model

$$y = a + b \, x + \varepsilon \tag{6}$$

where $y = \log(X)$, $a = \log(K_F \, c_{ref})$, $b = N$ and $x = \log(c_e/c_{ref})$ and *ε* is the error term, which is assumed to be normally distributed with zero mean. The WLS procedure minimizes the following $\chi^2$ merit function:

$$\chi^2 = \sum_{\lambda=1}^{L} \left( \frac{y_\lambda - a - N \, x_\lambda}{\sigma_{\varepsilon,\lambda}} \right)^2 \tag{7}$$

where *λ* is the index for the concentration level, *L* is the number of concentration levels, $y_\lambda$ and $x_\lambda$ are the logarithms of *X* and $c_e/c_{ref}$, respectively, and $\sigma_{\varepsilon,\lambda}$ is the standard deviation of *ε* at concentration level *λ*.

The approach of the 'effective variance' advocated by Tellinghuisen & Bolster (2010) and Bolster & Tellinghuisen (2010), is based on summing the direct and indirect contributions to the variance in $X$ resulting from uncertainty in $c_e$, as illustrated in Figure 1 (admittedly, this figure assumes an unrealistically high error in $c_e$ for demonstration purposes; furthermore this figure assumes no error in $c_i$, which would have led to displacement of the $X_d$ line). Sorption experiments are usually carried out for a range of $c_i$ levels and a few replicates sorption systems for each level $c_i$. Then, all replicates have the same $c_i$. Using the approach of the effective variance and assuming CVs of $c_e$ and $c_i$ of less than 10%, one can derive (see the Annex) that for such experiments,

$$\sigma_\varepsilon = \frac{1}{\delta \ln(10)} \sqrt{\gamma_i^2 + \gamma_e^2 \frac{[1 - \delta(1 - N)]^2}{U}} \tag{8}$$

where $\sigma_\varepsilon$ is the standard deviation of $\varepsilon$, $\gamma_i$ and $\gamma_e$ are the CVs of $c_i$ and $c_e$, respectively, and $U$ is the number of replicate $c_e$ measurements. The derivation indicates that the fit must be based on the average $\log(X)$ and $\log(c_e/c_{ref})$ values of the replicates because the errors in the replicate sorption points are correlated, as all replicates are based on the same $c_i$. In case of experiments with separate $c_i$ measurements for each replicate, no averaging of sorption points should be performed and Equations 5 and 8 can be used with $U = 1$ while setting $L$ equal to the total number of sorption points. Using Equation 8, it is thus possible to generate quotients $\varepsilon/\sigma_\varepsilon$ that are standard normally distributed, which is a prerequisite for the $\chi^2$ merit function of Equation 7 to follow the corresponding $\chi^2$ distribution (as shown in the Annex) and thus for obtaining correct CVs for the estimated parameters $K_F$ and $N$ of Equation 5.

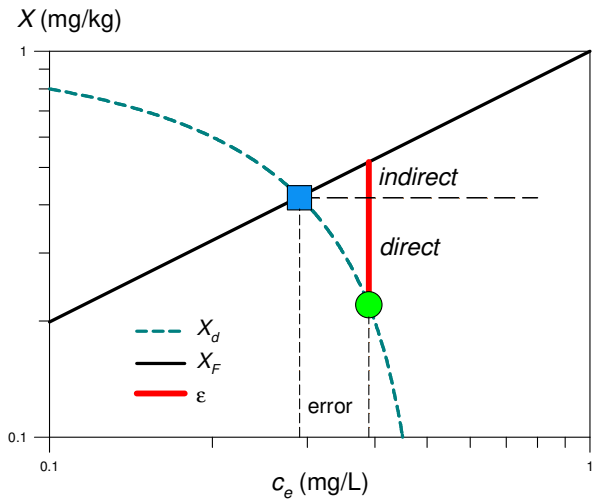

**Figure 1.** Illustration of the contributions of the direct and indirect variance to the variance of $\varepsilon$ resulting from an error in $c_e$. The $X_F$ line is Equation 1 with $K_F = 1$ L/kg, $N = 0.7$, and $c_{ref} = 1$ mg/L, and the $X_d$ line is Equation 2 with $c_i = 0.5$ mg/L and $V/M = 2$ L/kg. The square is the true point of the sorption isotherm, and the circle is the measured sorption point. It is assumed that there is no error in $c_i$.

The standard deviations of $a$ and $N$ ($\sigma_a$ and $\sigma_N$) were estimated following Press et al.(2007):

$$\sigma_a^2 = \frac{1}{\sum w_\lambda}\left(1 + \frac{\sum (w_\lambda x_\lambda)^2}{\sum w_\lambda \sum w_\lambda (x_\lambda - x_m)^2}\right) \quad (9)$$

$$\sigma_N^2 = \frac{1}{\sum w_\lambda (x_\lambda - x_m)^2} \quad (10)$$

$$w_\lambda \equiv \frac{1}{\sigma_{\varepsilon,\lambda}^2} \quad (11)$$

$$x_m \equiv \frac{\sum w_\lambda x_\lambda}{\sum w_\lambda} \quad (12)$$

where $w_\lambda$ is the weight at concentration level $\lambda$ and $x_m$ is the weighted average of the $x_\lambda$ values. Equations 9 to 12 assume that the weights are adequate (i.e., the a priori estimation procedure; Bolster & Tellinghuisen, 2010; Tellinghuisen, 2000) and thus do not use information on the differences between fitted and measured values (no $y$ values appear in Equations 9 to 12).

Using error propagation theory, the standard deviation of $K_F$ can be derived from $\sigma_a$ with

$$\sigma_{K_F} = \ln(10)\, K_F\, \sigma_a \quad (13)$$

To obtain insight into the $\sigma_\varepsilon$ function of Equation 8, we consider the $[1 - \delta(1-N)]^2$ term, which determines the contribution of the $\gamma_e^2$ term to $\sigma_\varepsilon$ (see Equation 8). Figure 2A shows that $[1 - \delta(1-N)]^2$ ranges between 0 and 1 and continuously decreases with increasing $\delta$ and decreasing $N$. Thus, $\gamma_e$ is hardly relevant for the estimation of $\sigma_\varepsilon$ if $\delta$ is high and $N$ is small, such as in systems with strong sorption and strongly curved isotherms, e.g. typical for phosphate sorption (e.g. Bolster & Tellinghuisen, 2010). Next, we analyze the full $\sigma_\varepsilon$ function (Equation 8) by considering the quotient $\sigma_\varepsilon/\gamma_e$ for $\gamma_i/\gamma_e = 0.5$ in Figure 2B (assuming that $\gamma_i$ is no more than half of $\gamma_e$; Tellinghuisen & Bolster, 2010). This quotient $\sigma_\varepsilon/\gamma_e$ increases considerably with decreasing $\delta$ and Equation 8 shows that it becomes infinite for $\delta = 0$. The reason for this high $\sigma_\varepsilon$ at low $\delta$ values is that $X$ is derived from the difference between $c_i$ and $c_e$ (Equation 2); thus, if $c_i$ and $c_e$ are almost equal (i.e., $\delta$ is nearly zero), the error in $X$ becomes extremely large (Boesten, 2015), and thus, $\sigma_\varepsilon$ becomes extremely large. Setting $\gamma_i$ to zero had a comparatively small effect on $\sigma_\varepsilon/\gamma_e$ for most of the $\delta$-$N$ plane (Figure 2B). The effect of setting $\gamma_i$ to zero only becomes considerable in Figure 2B when $\delta$ approaches 100% and $N$ becomes smaller than approximately 0.5. Thus, this confirms the conclusion from Figure 2A that the $\gamma_e^2$ term is relatively unimportant at high $\delta$ and small $N$. The calculation in Figure 2B for $\gamma_i/\gamma_e = 0.5$ was for three replicates ($U = 3$). This calculation was repeated for $U = 5$ (a rather high number of replicates) but this change had only a small effect on $\sigma_\varepsilon/\gamma_e$: the $\sigma_\varepsilon/\gamma_e$ for $U = 5$ was always between 0.88 and 1.0 times the $\sigma_\varepsilon/\gamma_e$ for $U = 3$ (data not shown).

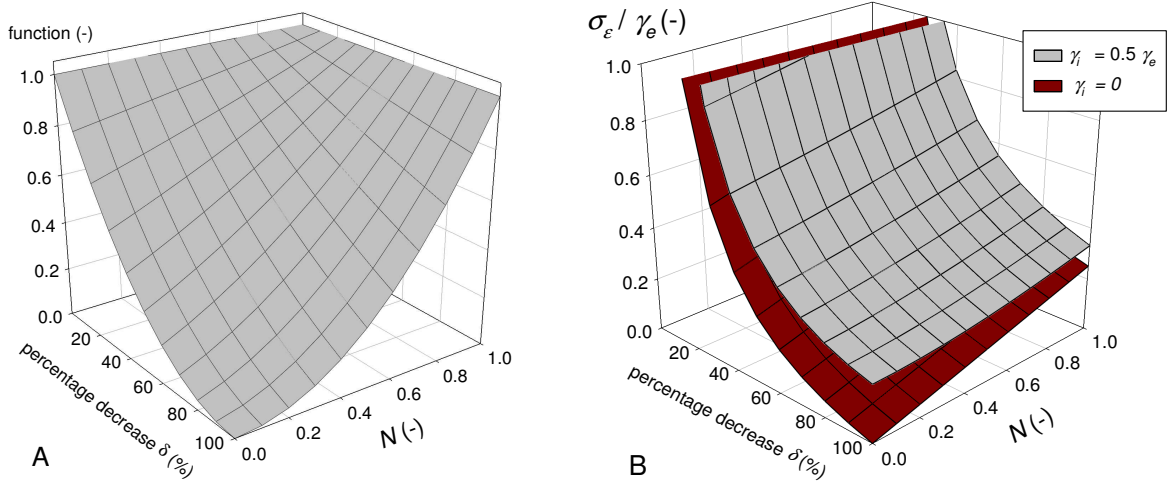

**Figure 2.** The function $[1- \delta(1-N)]^2$ (part A) and the quotient $\sigma_\varepsilon/\gamma_e$ (part B) as a function of the percentage decrease $\delta$ (%) and of $N$. The quotient $\sigma_\varepsilon/\gamma_e$ was calculated with Equation 8 assuming that $U = 3$ and either $\gamma_i = 0.5\ \gamma_e$ or $\gamma_i = 0$, as indicated.

The strong dependency of $\sigma_\varepsilon$ on $\delta$, as shown in Figure 2B, indicates that there may be considerable differences between the $\sigma_\varepsilon$ values of the sorption points (i.e., strong heteroscedasticity; Bolster & Tellinghuisen, 2010) at the different concentration levels of a single adsorption isotherm in cases where $\delta$ varies considerably between these concentration levels. Large variations of $\delta$ between concentration levels can be expected at moderate $K_F$ values in combination with low $N$ values. Differences between $\delta$ values are small at low $K_F$ values because all $\delta$ values are then close to zero. These differences are also small at high $K_F$ values because all $\delta$ values then approach 100%. These differences are also small at high $N$ values because they converge to zero when $N$ approaches 1.

# 3 Methods

## 3.1 Procedures of the Monte Carlo simulations

The simulations mimicked the following experimental procedure: (i) aqueous solutions of the sorbate are prepared for a range of $c_i$ levels, (ii) a known volume of these solutions is added to sorbent samples of known mass (triplicate systems for each concentration level), (iii) the suspension is equilibrated, and then, $c_e$ is measured, (vi) $X$ is derived from the difference between the added sorbate mass and the sorbate mass in the liquid phase after equilibration (Equation 2).

Monte-Carlo simulations were carried out for sorption systems with a wide range of Freundlich parameters. $K_F$ varied between 0.5 and 10 L/kg in 25 steps (i.e., 26 values) that changed by a constant factor while keeping the sorbent-liquid ratio $R$ fixed at 1 kg/L. It follows from Equations 4 and 8 that systems with the same dimensionless $R\,K_F$ but different $R$ and $K_F$ values behave in the same manner. This conclusion was confirmed by Monte-Carlo simulations, thus, this procedure can be considered as varying $R\,K_F$ between 0.5 and 10. The exponent $N$ varied between 0.2 and 1 in 20 steps (i.e., 21 values, giving $21 \times 26 = 546$ systems). The range of $N$ was limited to values below 1 because $N$ values above 1 occur only for a small fraction of the relevant sorbate-sorbent systems (e.g. for the sorption of monovalent cations to clay minerals in the presence of divalent cations, generating 'favorable exchange' isotherms; Bolt & Bruggenwert, 1976; Gaillardon et al., 1977). Systems that resulted in $\delta$ values below 30% were discarded (i.e., 121 systems, leaving 425 systems) because the accuracy of the estimated $K_F$ and $N$ from such systems is low (Boesten, 2015). Thus, such studies must be considered inadequate, and therefore, it is not interesting to analyze such systems. Five $c_i$ levels were assumed ($L = 5$): 0.1, 0.32, 1, 3.2 and 10 mg/L (these values differ by a factor of $10^{1/2}$ from each other; thus, they are equidistant on a logarithmic scale). There were three replicate sorption systems ($U = 3$) per concentration level and only a single measurement of $c_i$ at each concentration level.

The errors in $c_i$ and $c_e$ were assumed to be independent and normally distributed around their true values with the same CV for all concentration levels, with values of 1% for $\gamma_i$ and 5% for $\gamma_e$. The variability in the initial concentrations ($\gamma_i$) is only caused by the chemical analysis of the water, whereas the variability in the equilibrium concentrations ($\gamma_e$) is also determined by differences in sorption between replicate systems. A $\gamma_e$ of 5% is a realistic value in the high range for sorption studies with pesticides when measuring concentrations with high pressure liquid chromatography or gas-liquid chromatography and it is also considered realistic for phosphate (Bolster & Tellinghuisen, 2010; Boesten, 2015).

For each Freundlich sorption system, 10,000 isotherms were generated (thus simulating 10,000 sorption experiments). For each isotherm, five $c_i$ values (one for each concentration level) and 15 $c_e$ values (because of the three replicates) were drawn. Thus, 15 $c_e - X$ pairs were generated. From the three replicate $c_e$ and $X$ values at each concentration level, the averages of the logarithms of $c_e/c_{ref}$ and $X$ were calculated, which were fitted to Equation 5. From each fit the CVs of $K_F$ and $N$ were estimated with Eqns 13 and 10, respectively.

### 3.2 Description of the simulation cases *I* to *IV*

For all simulation cases, the first step was a fit with $\sigma_\varepsilon = 1$ (i.e., using ULS), resulting in estimates of $K_F$ and $N$. For the cases *I* and *II*, the second step was a fit with WLS based on $\sigma_\varepsilon$ values derived from Equation 8 for the five concentration levels (using the $N$ value of the first fit) but using different calculation procedures for $\sigma_\varepsilon$. In the first set of simulations (case *I*), $\sigma_\varepsilon$ was calculated for each concentration level with Equation 8 using the true values of $\delta$, $\gamma_i$ and $\gamma_e$. This approach was used to demonstrate the accuracy of Equation 8 by comparing the 10,000 CVs of $K_F$ and $N$ estimated by the WLS model for each of the generated isotherms with the estimated true CVs based on the populations of the 10,000 fitted $K_F$ and $N$ values.

Use of the true values of $\delta$, $\gamma_i$ and $\gamma_e$ is appropriate when testing the consistency of concepts in the case *I* simulations but is of little value in experimental practice because the true values of $\delta$, $\gamma_i$ and $\gamma_e$ are unknown. Thus, in case *II*, $\sigma_\varepsilon$ was again based on Equation 8 for the five concentration levels but now, $\delta$, $\gamma_i$ and $\gamma_e$ were estimated from the data of each generated isotherm. The purpose of these case *II* simulations was to demonstrate that use of Equation 8 leads to reasonably accurate CVs of $K_F$ and $N$ for individual sorption isotherms. For each concentration level, $\delta$ was calculated as the average of the triplicate measured $\delta$ values. The calculation of $\gamma_e$ was based the sample variance, $s^2$, of each set of triplicate $c_e$ measurements:

$$s^2 = \frac{1}{U-1} \sum_{u=1}^{U} (c_{e,u} - c_{e,av})^2 \qquad (14)$$

where $c_{e,av}$ is the average $c_e$ of the triplicates. The $\gamma_e^2$ at each concentration level was calculated as $s^2/(c_{e,av})^2$, and thereafter, the average $\gamma_e^2$ of all concentration levels was calculated and used in Equation 8. For $\gamma_i$, this procedure was impossible because there are usually no replicate $c_i$ measurements at a certain concentration level. However, the expected concentrations at each concentration level are typically known because of the experimental design of the study. Thus, $\gamma_i$ was estimated by (i) calculating the quotient of each measured $c_i$ value divided by its expected value, (ii) using the average of these quotients to calculate revised quotients of measured $c_i$ values divided by the expected $c_i$ values at all concentration levels, and (iii) calculating the sample standard deviation of these quotients (similar to Equation 14). This complex procedure is clarified by the following example. Assume that there are two concentration levels with expected values of 1.0 and 10.0 mg/L and that the two measured $c_i$ values were 0.94 and 9.8 mg/L. The two quotients are then 0.94 and 0.98 (step i), so the average quotient is 0.96. This gives revised quotients of 0.94/0.96 = 0.98 and 0.98/0.96 = 1.02 (step ii). Calculating the sample standard deviation of 0.98 and 1.02 (step i) yields $\gamma_i$ = 2.8%. Monte Carlo simulations confirmed that these procedures for estimating $\gamma_i$ and $\gamma_e$ generated the expected distributions of the $\gamma_e^2$ and $\gamma_i^2$ values for the degrees of freedom $L$-1 for $\gamma_i^2$ and $L(U$-1) for $\gamma_e^2$ (the expected distributions are given by the corresponding reduced $\chi^2$ distributions defined by Tellinghuisen, 2000).

In the next set of simulations (case *III*), the standard ULS procedure was used to serve as a benchmark for the improvement that can be achieved with Equation 8. The first step of the procedure was a fit with ULS as for all cases. The estimated variances obtained from Equations 9 and 10 were then multiplied by the factor $F_{ULS}$ to obtain a posteriori estimates of these variances (Bolster & Tellinghuisen, 2010):

$$F_{ULS} = \frac{\sum_{\lambda=1}^{L} (y_\lambda - a - N x_\lambda)^2}{\nu} \tag{15}$$

where $\nu$ is the degrees of freedom (i.e., $L$-2). As Equation 15 does not make any use of the knowledge of the structure of the errors, the extent to which using Equation 8 leads to improvement of the estimated CVs of $K_F$ and $N$ can be determined by comparing results of case *II* with those of case *III*.

Case *IV* considered another alternative, i.e. weights that are known only in a relative sense. The first step of the procedure was a fit with ULS as for all cases. The second step was a fit with WLS using $\sigma_\varepsilon$ values that were 0.1 times the $\sigma_\varepsilon$ values of case *I* (i.e., using true values of $\delta$, $\gamma_i$ and $\gamma_e$, which is the most favorable case possible for case *IV*). This 0.1 is an arbitrary multiplication factor (such that the absolute values of the weights increased by a factor of 100; see Equation 12). The estimated variances obtained from Equations 9 and 10 were then multiplied with the following factor $F_{WLS}$ to obtain the a posteriori estimates of the parameter variances (Bolster & Tellinghuisen, 2010):

$$F_{WLS} = \frac{\sum_{\lambda=1}^{L} w_\lambda (y_\lambda - a - N x_\lambda)^2}{\nu} \tag{16}$$

# 4 Results and discussion

## 4.1 Simulated $K_F$ and $N$ values and their true CVs

The average $K_F$ and $N$ values as estimated from the fitted 10,000 values of $K_F$ and $N$ of each sorption system, never differed from the true values by more than 0.4% for all cases (*I* to *IV*). The true CVs of $K_F$ and $N$ were estimated from the 10,000 values of $K_F$ and $N$ found for case *I* because the weights from case *I* are the most reliable. These true CVs were closely related to the $R\ K_F$ of the system. The CV of $K_F$ decreased from approximately 4% at $R\ K_F = 0.5$ to 1-2% at $R\ K_F = 10$. The CVs of $N$ for these systems decreased from approximately 2.5% at $R\ K_F = 0.5$ to 0.5-1% at $R\ K_F = 10$. These CVs are quite low, indicating that the simulated studies had an adequate experimental design for determining these parameters. The true CVs of each sorption system will be used as the yardstick for the accuracy of the CVs estimated for the individual isotherms of the different cases.

## 4.2 Accuracy of the CVs estimated using the true error parameters (Case *I*)

The results for the case *I* simulations in Figures 3A and 3B show that the CVs derived from the 10,000 standard deviations of $K_F$ and $N$ (based on Eqns 13 and 10) corresponded well with the true CVs. The median CV derived from the 10,000 standard deviations was always nearly exactly equal to the true CV both for $K_F$ and $N$. The 2.5$^{th}$ and 97.5$^{th}$ percentiles of the CVs of $N$ were also always nearly exactly equal to the true CVs. The 2.5$^{th}$ and 97.5$^{th}$ percentiles of the CVs of $K_F$ were rather close to the true CVs for $R\ K_F > 1$. However, at $R\ K_F = 0.5$ they differed from the true CVs by approximately 10%. The reason for this discrepancy at $R\ K_F = 0.5$ is likely that the estimation of the $K_F$ and $N$ becomes cumbersome when $\delta$ (and thus $R\ K_F$; see Equation 4) becomes small. Thus, Figures 3A and 3B indicate that the standard deviations of $K_F$ and $N$ based on the $\sigma_\varepsilon$ values from Equation 8 are close to the correct values when the true values of $\delta$, $\gamma_i$ and $\gamma_e$ are used. Figures 3A and 3B show that the differences between the values of a certain percentile at a certain $R\ K_F$ are very small. As Figures 3A and 3B contain results for $N$ values ranging from 0.2 to 1.0, this indicates that $N$ had a small effect on the CVs. The differences between the $K_F$ and $N$ values derived from the first and second fit were small. The ratio of the values from the first fit divided by those of the second fit was calculated (both the average and standard deviation of the 10,000 ratios) for each of the 425 sorption systems: for $N$, the average ranged from 0.9987 to 1.0001, and the standard deviation ranged from 0.0000 to 0.0156. For $K_F$, these ranges were from 0.9979 to 1.0002 and from 0.0000 to 0.0291, respectively.

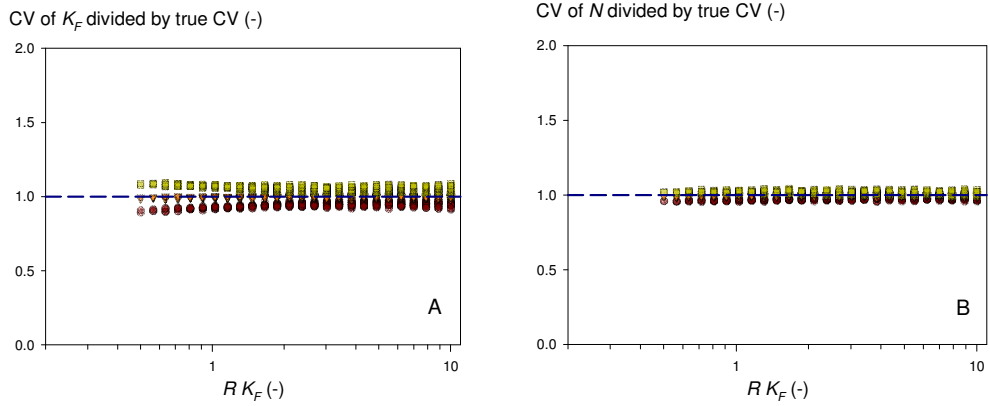

**Figure 3.** Ratios of CVs of $K_F$ or $N$ derived from fits to single sorption isotherms divided by their true CV as a function of $R\,K_F$ for case *I*, i.e. using true parameters for Equation 8. Circles, triangles and squares are 2.5$^{th}$, 50$^{th}$ and 97.5$^{th}$ percentiles of the ratios of the population of 10,000 values calculated for each sorption system, respectively.

## 4.3 Accuracy of the CVs estimated using the estimated error parameters (Case *II*)

The CVs of $K_F$ and $N$ found for the case *II* simulations are presented in Figures 4A and 4B as a function of $N$ because they showed more variation with $N$ than with $K_F$, as is illustrated by the narrow bands of points in these figures (which include all $K_F$ values). The median of the estimated CVs corresponded well with the true CVs. However, the 2.5$^{th}$ and 97.5$^{th}$ percentiles of the CVs differ from the true CVs by 50-60% at $N$ values below 0.4 and by approximately 40% for higher $N$ values. The only difference between Figures 3A and 3 B and Figures 4A and 4B is the procedure used to obtain the $\delta$, $\gamma_i$ and $\gamma_e$ values (true versus estimated, respectively). Thus, the estimated combinations of $\delta$, $\gamma_i$ and $\gamma_e$ values must be inaccurate. Furthermore, the differences between the $K_F$ and $N$ values derived from the first and second fits were small (as in Case *I*). For $N$, the average ratio (as described in the previous paragraph) ranged from 0.9974 to 1.0000, and the standard deviation ranged from 0.0001 to 0.0155. For $K_F$, these ranges were from 0.9932 to 1.0000 and from 0.0002 to 0.0289, respectively. Additional simulations were carried out to determine the extent to which the uncertainty in each of the three parameters $\delta$, $\gamma_i$ and $\gamma_e$ contributed to the uncertainty in the estimated CVs as found for case *II*. This was done by using estimated values of one of the three parameters in combination with true values of the other two parameters. Using estimated values of $\delta$ in combination with true values of $\gamma_i$ and $\gamma_e$ gave results similar to Figures 3A and 3B, indicating that the estimation of $\delta$ was sufficiently accurate (data not shown). Results obtained with estimated values of $\gamma_i$ or $\gamma_e$ showed that the estimations of $\gamma_e$ and $\gamma_i$ were the most important causes of the differences between the cases *I* and *II* in the high and low range of $N$ values, respectively (data not shown). The effect of the estimated $\gamma_e$ values is consistent with the term $[1- \delta(1-N)]^2$ of Equation 8 approaching zero in Figure 2B for low $N$ values (and thus leading to a small contribution of $\gamma_e$ to $\sigma_\varepsilon$) and the effect of the estimated $\gamma_i$ values is consistent with the small difference between the two planes at high $N$ values in Figure 2B, leading to a small contribution of $\gamma_i$ to $\sigma_\varepsilon$ at high $N$ values.

The reason for the small effect of estimating $\delta$ is the rather high accuracy of the estimated $\delta$ for $\delta > 30\%$. Using error propagation theory, it can be derived from Equation 3 that

$$\gamma_\delta = \left(\frac{1-\delta}{\delta}\right)\sqrt{\gamma_i^2 + \frac{\gamma_e^2}{U}} \tag{17}$$

where $\gamma_\delta$ is the CV of the estimated $\delta$. Thus, if, for instance, $\delta$ equals 50% (i.e., 0.5), then $\gamma_\delta$ equals 3.1% for $\gamma_e = 5\%$, $\gamma_i = 1\%$, and $U = 3$. Thus, the background of the relatively small effect of estimating $\delta$ is that estimating an average from a few replicates is more accurate than estimating a variance from a few replicates.

The estimation of $\gamma_i$ can be improved relatively easily because it is only determined by the analytical procedure, whereas $\gamma_e$ is also influenced by the variability of the sorption system (i.e. the mass of sorbent and the homogenization process of the sorbent in the case of heterogeneous sorbents, such as soils). Thus, if for instance, a number of sorption isotherms are measured on the same day using the same aqueous solutions for different concentration levels, data on the measured initial concentrations from different isotherms could be combined to improve the estimated $\gamma_i$. As follows from the above, this approach will only help in the case of low $N$ values because $\gamma_i$ contributes only slightly to $\sigma_\varepsilon$ for high $N$ values. More sophisticated approaches for estimating the variance functions could be tested (Tellinghuisen & Bolster, 2010), but such approaches are specific for a certain type of sorbate-sorbent systems and thus beyond the scope of this work.

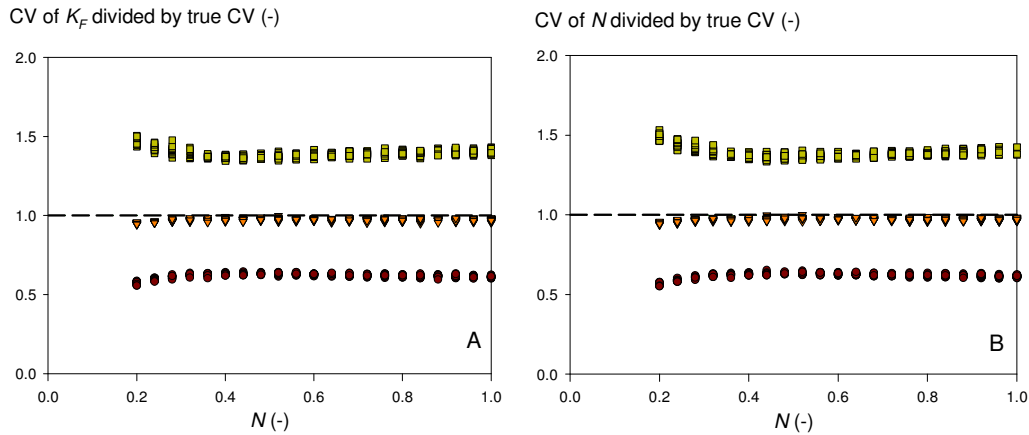

**Figure 4**. Ratios of CVs of $K_F$ or $N$ derived from fits to single sorption isotherms divided by their true CV a function of $N$ for case *II*, i.e. using estimated parameters for Equation 8. Circles, triangles and squares are 2.5[th], 50[th] and 97.5[th] percentiles of the ratios of the population of 10,000 values calculated for each sorption system, respectively.

## 4.4 Accuracy of the estimated CVs for ULS (Case *III*)

The results for case *III* simulations in Figures 5A and 5B show that the median CVs of $K_F$ and $N$ generated by the conventional ULS procedure were mostly lower than the true CVs. Furthermore, the 2.5[th] and 97.5[th] percentiles of these CVs differed considerably more from their true values than those found in Figures 4A and 4B when using Equation 8. Part of the 97.5[th]-percentile data are not shown because they were larger than the maximum of the vertical axis (which was kept at 2.0 for consistency with Figures 4A and 4B). Further analysis revealed that there is a close relationship between the ratio of the 97.5[th] percentile CVs of $K_F$ or $N$ divided by their true values (i.e., the data points shown in Figures 5A and 5B) and the difference in $\delta$ between the highest and the lowest concentration levels of an isotherm: as this difference in $\delta$ increases, this ratio becomes larger (this difference is largest for combinations of small $R$ $K_F$ and small $N$ values). This effect of the difference in $\delta$ is understandable because for large differences in $\delta$, Equation 8 leads to large differences between the $\sigma_\varepsilon$ values of the different concentration levels (i.e., strong heteroscedasticity), whereas ULS uses the same $\sigma_\varepsilon$ for all concentration levels.

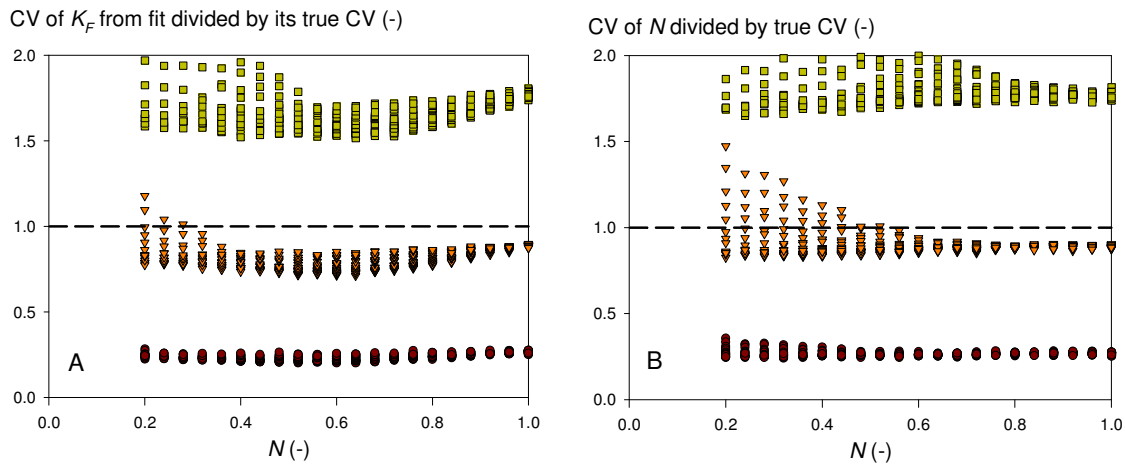

**Figure 5.** Ratios of CVs of $K_F$ or $N$ derived from fits to single sorption isotherms divided by their true CV as a function of $N$ for case *III* (ULS with Equation 15). Circles, triangles and squares are 2.5[th], 50[th] and 97.5[th] percentiles of the ratios of the population of 10,000 values calculated for each sorption system, respectively; 97.5[th] percentiles points with a ratio larger than 2 are not shown; the numbers of ratios larger than 2 corresponded with 6% and 16% of the total number of points for the 97.5[th] percentiles for the CV of $K_F$ and $N$, respectively (the maximum ratio of these points was 3.5 for $K_F$ and 4.5 for $N$).

## 4.5 Accuracy of the estimated CVs for relative weights (Case *IV*)

Comparison of Figures 6A and 6B with Figures 4A and 4B shows that using perfectly correct weights on a relative scale leads to a posteriori CVs that are worse than those found with $\delta$, $\gamma_i$ and $\gamma_e$ values that were estimated from the isotherm data (case *II*). Thus, using inaccurate weights in combination with the a priori approach of case *II* gives better CVs than using perfect relative weights in combination with the a posteriori approach of case *IV*. Thus, the a priori procedure is considerably better than the a posteriori procedure. Using correct relative weights (case *IV*) gives better results than the ULS procedure (compare Figures 5A-5B with Figures 6A-6B).

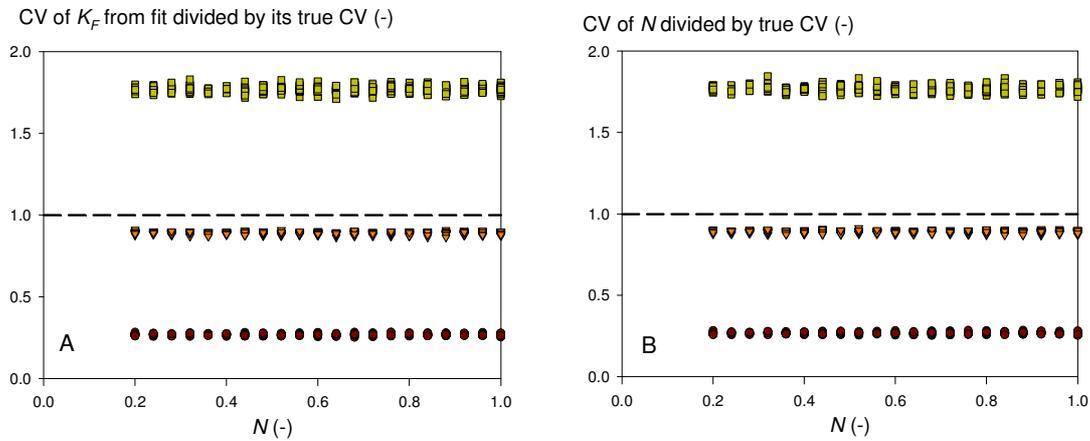

**Figure 6.** Ratios of CVs of $K_F$ or $N$ derived from fits to single sorption isotherms divided by their true CV as a function of $N$ for case *IV* (relative weights with Equation 16). Circles, triangles and squares are 2.5[th], 50[th] and 97.5[th] percentiles of the ratios of the population of 10,000 values calculated for each sorption system, respectively.

# 5   Conclusions

The derived analytical approximation for the weights to be used in the WLS analysis resulted in the correct $\chi^2$ distribution of the sum of squares. The weights were found to decrease with increasing concentration, increasing Freundlich exponent, and increasing CVs of the initial and equilibrium concentrations. The analytical approximation improved the accuracy of the CVs of the Freundlich parameters considerably when compared to ULS. Thus this work provides a comparatively simple method (i.e. use Equations 8, 9, 10 and 13) for obtaining better estimates of the CVs of the Freundlich parameters $K_F$ and $N$. It is recommended to use this method whenever these CVs are relevant for the further use and interpretation of the estimated Freundlich parameters.

**Acknowlegdements**

I thank Leo Boumans for his many useful suggestions and my son Frank for providing Equation A20.

**References**

Barrow, N.J. 2008. The description of sorption curves. European Journal of Soil Science 59, 900-910.

Boesten, J.J.T.I. & van der Pas, L.J.T. 2000. Movement of water, bromide and the pesticides ethoprophos and bentazone in a sandy soil: the Vredepeel data set. Agricultural Water Management, 44, 21-42.

Boesten, J.J.T.I. 2007. Simulation of pesticide leaching in the field and in zero-tension lysimeters. Vadose Zone Journal 6, 793-804.

Boesten, J.J.T.I. 2015. Effects of random and systematic errors on Freundlich parameters for pesticide sorption. Soil Sci. Soc. Am. J. 79, 1306-1318.

Bolster, C.H. & Tellinghuisen, J. 2010. On the significance of properly weighting sorption data for least square analysis. Soil Sci. Soc. Am. J. 74, 670-679.

Bolt, G.H. & Bruggenwert, M.G.M. 1976. Soil chemistry. A. Basic elements, Elsevier, Amsterdam.

Bowman, B.T. 1982. Conversion of Freundlich adsorption K values to the mole fraction format and the use of Sy values to express relative adsorption of pesticides. Soil. Sci. Soc. Am. J. 46, 740-743.

Buekers, J., Amery, F., Maes, A. & Smolders, E. 2008. Long-term reactions of Ni, Zn and Cd with iron oxyhydroxides depend on crystallinity and structure and on metal concentrations. European Journal of Soil Science, 59, 706-715.

Che, M., Loux, M.M., Traina, S.J. & Logan, T.J. 1992. Effect of pH on sorption and desorption of imazaquin and imazethapyr on clays and humic acid. J. Environ. Qual. 21, 698-703.

Draper, N.R. & Smith, H. 1998. Applied regression analysis. Third edition, Wiley, New York.

Freundlich, H. 1907. Über die Adsorption in Lösungen. Zeitschrift für physikalische Chemie, Stöchiometrie und Verwandtschaftslehre 57, 385–470.

Gaillardon, P., Calvet, R. & Tercé, M. 1977. Adsorption et desorption de la terbutryne par une montmorillonite-Ca et des acides humiques seuls ou en melanges. Weed Research 17, 41-48.

Georgievics, G. & Löwy, E. 1895. Űber das Wesen des Färbeprocesses. Vertheilung von Methylenblau zwischen Wasser un mercerisirter Cellulose. Monatshefte für Chemie 16, 345-350.

He, Y., Zhang, J., Wang, H.Z., Shi, J.C. & Xu, J.M. 2011. Can assessing for potential contribution of soil organic and inorganic components for butachlor sorption be improved ? J. Environ. Qual. 40, 1705-1713.

Jones, R.G. & Loeppert, R.H. 2013. Calcite surface adsorption of As(V), As(III), MMAs(V), and DMAs(V) and the impact of calcium and phosphate. Soil Science Society of America Journal, 77, 83-93.

Kadyampakeni, D.M., Morgan, K.T., Mahmoud, K., Schumann, A. & Nkedi-Kizza, P. 2014. Phosphorus and potassium distribution and adsorption on two Florida sandy soils. Soil Science Society of America Journal, 78, 325-334.

Kinniburgh, D.G., Barker, J.A. & Whitfield, M. 1983. A comparison of some simple adsorption isotherms for describing divalent cation adsorption by ferrihydrite. Journal of Colloid and Interface Science 95, 370-384.

Lado, M., Borisover, M. & Paz Gonzalez, A. 2013. Multifractal analysis of nitrogen adsorption isotherms obtained from organoclays exposed to different temperatures. Vadose Zone Journal, 12, 3, doi:10.2136/vzj2012.0206.

Larsbo, M., Stenstrom, J., Etana, A., Borjesson, E.& Jarvis, N.J. 2009. Herbicide sorption, degradation, and leaching in three Swedish soils under long-term conventional and reduced tillage. Soil & Tillage Research, 105, 200-208.

Mouni, L., Merabet, D., Robert, D. & Bouzaza, A. 2009. Batch studies for the investigation of the sorption of the heavy metals Pb2+ and Zn2+ onto Amizour soil (Algeria). Geoderma 154, 30-35.

Press, W.H., Teukolsky, S.A., Vetterling, W.T. & Flannery, B.P. 2007. Numerical recipes: the art of scientific computing, Cambridge University Press, New York.

Pronk, G.J., Heister, K., Woche, S.K., Totschke, K.U. & Koegel-Knabner, I. 2013. The phenanthrene-sorptive interface of an arable topsoil and its particle size fractions. European Journal of Soil Science, 64, 121-130.

Schmidt, G.C. 1894. Űber Adsorption. Zeitschrift für physikalische Chemie, Stöchiometrie und Verwandtschaftslehre 15, 56-64.

Sips, R. 1950. On the structure of a catalyst surface. II. Journal of Chemical Physics 18, 1024-1026.

Stougaard, R.N., Shea, P.J., Martin, A.R. 1990. Effect of soil type and pH on adsorption, mobility and efficacy of imazaquin and imazethapyr. Weed Science, 38, 67-73.

Tellinghuisen, J. A. 2000. Monte Carlo study of precision, bias, inconsistency, and non-gaussian distributions in nonlinear least squares. J. Phys. Chem A 104, 2834-2844.

Tellinghuisen, J. & Bolster, C.H. 2010. Least-squares analysis of phosphorus soil data with weighting from variance function estimation: a statistical case for the Freundlich isotherm. Environ. Sci. Technology 44, 5029-5034.

Toor, G.S. & Sims, J.T. 2015. Managing phosphorus leaching in mid-atlantic soils: importance of legacy sources. Vadose Zone Journal, 14, 12, doi:10.2136/vzj2015.08.0108.

Travis, C.C. & Etnier, E.L. 1981. A survey of sorption relationships for reactive solutes in soil. J. Environ. Qual. 10, 8-17.

Vega, F.A., Covelo, E.F. & Andrade, M.L. 2008. A versatile parameter for comparing the capacities of soils for sorption and retention of heavy metals dumped individually or together: Results for cadmium, copper and lead in twenty soil horizons. Journal of Colloid and Interface Science, 237, 275-286.

Van Riemsdijk, W.H., Bolt, G.H., Koopal, L.K. & Blaakmeer, J. 1986. Electrolyte adsorption on heterogeneous surfaces: adsorption models. Journal of Colloid and Interface Science 109, 129-228.

# Annex

**Derivation of an analytical expression for the standard deviation of the error in the Freundlich least squares analysis**

The approach of the 'effective variance' advocated by Tellinghuisen & Bolster (2010) and Bolster & Tellinghuisen (2010), is based on summing the direct and indirect contributions to the variance in *X* resulting from uncertainty in $c_e$, as illustrated in Figure 1 of the main paper. Bolster & Tellinghuisen (2010) calculate this effective variance using the approximation procedure derived by Clutton-Brock (1967). Figure 1 indicates that there is an alternative to this calculation procedure, i.e., considering all components of the standard deviation of the error term:

$$\varepsilon = \log(X) - a - N\log\left(\frac{c_e}{c_{ref}}\right) \tag{A1}$$

where $a \equiv \log(K_F\, c_{ref})$. Thus, there are two possible approaches to assess this effective variance. We will apply both approaches and we will show that these give indeed the same result.

We start with the approach based on Equation A1. Error propagation theory (e.g. Arfken & Weber, 2005) gives the following approximation for the variance $\sigma_u^2$ of a variable *u*, which is a function of two stochastic variables *v* and *w*:

$$\sigma_u^2 = \left(\frac{du}{dv}\right)^2 \sigma_v^2 + \left(\frac{du}{dw}\right)^2 \sigma_w^2 + 2\left(\frac{du}{dv}\right)\left(\frac{du}{dw}\right)\sigma(v,w) \tag{A2}$$

where $\sigma_v^2$ and $\sigma_w^2$ are the variances of *v* and *w* and $\sigma(v,w)$ is the covariance of *v* and *w*. Applying this to Equation A1, the following expression for the variance of $\varepsilon$, i.e., $\sigma_\varepsilon^2$ is obtained:

$$\sigma_\varepsilon^2 = \left(\frac{d\varepsilon}{d\log(X)}\right)^2 \sigma_{\log(X)}^2 + \left(\frac{d\varepsilon}{d\log(c_e)}\right)^2 \sigma_{\log(c_e)}^2 + 2\left(\frac{d\varepsilon}{d\log(X)}\right)\left(\frac{d\varepsilon}{d\log(c_e)}\right)\sigma(\log(X),\log(c_e)) \tag{A3}$$

$c_{ref}$ is omitted from Equation A3 because $\log(c_e/c_{ref}) = \log(c_e) - \log(c_{ref})$, so the variance of $\log(c_e/c_{ref})$ equals the variance of $\log(c_e)$ and $\sigma(\log(X), \log(c_e/c_{ref})) = \sigma(\log(X), \log(c_e))$. The first step is to obtain an expression for $\sigma^2_{\log(X)}$. Using error propagation theory, the following approximation of the coefficient of variation of *X* can be derived using Equations 2 and 3:

$$\gamma_X = \frac{1}{\delta}\sqrt{\gamma_i^2 + \gamma_e^2(1-\delta)^2} \tag{A4}$$

where $\gamma_X$, $\gamma_i$ and $\gamma_e$ are the coefficients of variation of $X$, $c_i$ and $c_e$, respectively. Using error propagation theory, it can also be derived that the standard deviation of the logarithm of a stochastic variable $\xi$, i.e., $\sigma_{\log \xi}$, can be approximated by

$$\sigma_{\log \xi} = \frac{\gamma_\xi}{\ln(10)} \tag{A5}$$

where $\gamma_\xi$ is the CV of $\xi$. This approximation is based on a Taylor series expansion using only the first derivative of $\log(\xi)$, truncating higher-order terms. Monte Carlo simulations showed that this approximation is accurate up to CVs of 0.10 (i.e., 10%). Using Equation A5 for $\gamma_X$ in Equation A4 yields:

$$\sigma_{\log X} = \frac{1}{\delta \ln(10)} \sqrt{\gamma_i^2 + \gamma_e^2 (1-\delta)^2} \tag{A6}$$

Thus, we still need an expression for $\sigma(\log(X), \log(c_e))$. Using Equation 2, this covariance can be written as

$$\sigma(\log(X), \log(c_e)) = \sigma\left(\log\left(\frac{V}{M}\right) + \log(c_i - c_e), \log(c_e)\right) \tag{A7}$$

Following the rules of the covariance, the constant $\log(V/M)$ has no effect on the covariance. Furthermore, the following approximation can be used:

$$\sigma(f, g) = \frac{df}{dg} \sigma(g, g) \tag{A8}$$

where $\sigma(g,g)$ is the variance of $g$ with $g = \log(c_e)$ and $f = \log(c_i - c_e)$. Thus, $df/dg$ can be written as

$$\frac{d \log(c_i - c_e)}{d \log(c_e)} = \frac{\dfrac{d \log(c_i - c_e)}{dc_e}}{\dfrac{d \log(c_e)}{dc_e}} = \frac{-c_e}{c_i - c_e} \tag{A9}$$

Using Equation 3 it can be shown that the right hand side of Equation A9 equals $(\delta - 1)/\delta$. Thus,

$$\sigma(\log(X), \log(c_e)) = \frac{\delta - 1}{\delta} \sigma^2_{\log(c_e)} \tag{A10}$$

As $\delta$ is smaller than 1 by definition, this covariance $\sigma(\log(X), \log(c_e))$ is always negative because a lower $c_e$ always leads to a higher $X$. When $\delta$ approaches 1 (i.e., nearly 100% sorption) the covariance (and thus also the correlation) becomes negligibly small because errors in $c_e$ hardly influence the error in $X$ (see Equation A6). When $\delta$ approaches 0 (i.e., nearly no sorption), the negative covariance becomes large because errors in $c_e$ have a large effect on the error in $X$.

Using Equation A10, Equation A3 becomes

$$\sigma_\varepsilon^2 = \sigma_{\log(X)}^2 + \sigma_{\log(c_e)}^2 \left( N^2 + 2N\frac{1-\delta}{\delta} \right) \tag{A11}$$

The term "$2N(1-\delta)/\delta$" is the covariance term of Equation A3. This term is positive so the negative $\sigma(\log(X),\log(c_e))$ leads to an increase of $\sigma_\varepsilon$. This increase is understandable from Equation A1 because $\varepsilon$ is calculated from the difference between $\log(X)$ and $\log(c_e)$. Using Equation A11 in combination with Equation A6 yields

$$\sigma_\varepsilon^2 = \left(\frac{1}{\delta \ln(10)}\right)^2 \left( \gamma_i^2 + \gamma_e^2 \left[1 - \delta(1-N)\right]^2 \right) \tag{A12}$$

With Equation A12, it is possible to generate quotients $\varepsilon/\sigma_\varepsilon$ that are standard normally distributed, which is a prerequisite for obtaining the correct $\chi^2$ distribution and thus for obtaining correct standard deviations for the estimated parameters.

If there are replicates, there is the additional problem of correlation between the $\chi^2$ variables of replicates because replicates are based on the same $c_i$ (via measurement of the same concentration in the added liquid, $c_a$). In the case of two replicates, we have pairs of

$$Z_1 = \log(X_1) - a - N \log\left(\frac{c_{e,1}}{c_{ref}}\right) \tag{A13}$$

$$Z_2 = \log(X_2) - a - N \log\left(\frac{c_{e,2}}{c_{ref}}\right) \tag{A14}$$

where $Z_1$ and $Z_2$ are correlated because they are based on the same measurement of $c_i$. It is assumed that the two sorption points are averaged before the fit, i.e., $\log(X_1)$ & $\log(X_2)$ and $\log(c_{e,1}/c_{ref})$ & $\log(c_{e,2}/c_{ref})$ are averaged, which means that $\varepsilon$ is defined as

$$\varepsilon \equiv \tfrac{1}{2} Z_1 + \tfrac{1}{2} Z_2 \tag{A15}$$

Error propagation (Equation A2) gives the following approximation for the variance of $\varepsilon$

$$\sigma_\varepsilon^2 = \left(\frac{d\varepsilon}{dZ_1}\right)^2 \sigma_{z_1}^2 + \left(\frac{d\varepsilon}{dZ_2}\right)^2 \sigma_{z_2}^2 + 2\left(\frac{d\varepsilon}{dZ_1}\right)\left(\frac{d\varepsilon}{dZ_2}\right) \sigma(Z_1, Z_2) = \tfrac{1}{4}\sigma_z^2 + \tfrac{1}{4}\sigma_z^2 + 2 \cdot \tfrac{1}{2} \cdot \tfrac{1}{2} \sigma(Z_1, Z_2) \tag{A16}$$

We know $\sigma_z$ (equal to the $\sigma_\varepsilon$ of Equation A12) but do not know the covariance $\sigma(Z_1, Z_2)$. Considering Equations A13 and A14, there is no correllation between the pairs $X_1$ - $c_{e,2}$, $X_2$ - $c_{e,1}$, and $c_{e,1}$ - $c_{e,2}$. Following the calculation rules for covariances, we know that

$\sigma(Z_1, Z_2) = \sigma(\log(X_1), \log(X_2))$ (A17)

Using Equation 2 yields

$$\sigma(\log(X_1), \log(X_2)) = \sigma\left( \log\left(\frac{V}{M}\right) + \log(c_i - c_{e,1}), \log\left(\frac{V}{M}\right) + \log(c_i - c_{e,2}) \right) \tag{A18}$$

and thus

$$\sigma(\log(X_1), \log(X_2)) = \sigma(\log(c_i - c_{e,1}), \log(c_i - c_{e,2}))  \quad (A19)$$

We use the approximation (Anonymous, 2013)

$$\sigma(\ln(F), \ln(G)) \approx \ln\left(1 + \frac{\sigma(F,G)}{E(F)\,E(G)}\right) \approx \frac{\sigma(F,G)}{F\,G} \quad (A20)$$

where $E(F)$ is the expected $F$ and $E(G)$ is the expected $G$.

Combining Equations A19 and A20 results in

$$\sigma(\log(X_1), \log(X_2)) = \frac{\sigma((c_i - c_{e,1}),(c_i - c_{e,2}))}{(\ln(10))^2 \; (c_i - c_{e,1})(c_i - c_{e,2})} \quad (A21)$$

Because there is no correlation between $c_i$ and $c_{e,1}$ or $c_{e,2}$ and between $c_{e,1}$ and $c_{e,2}$, it follows that

$$\sigma((c_i - c_{e,1}),(c_i - c_{e,2})) = \sigma(c_i, c_i) = \sigma_i^2 \quad (A22)$$

Using Equation 3, it can be shown that Equation A21 then results in

$$\sigma(\log(X_1), \log(X_2)) = \left(\frac{\gamma_i}{\delta \ln(10)}\right)^2 \quad (A23)$$

Combination of A12, A16 and A23 then yields

$$\sigma_\varepsilon = \frac{1}{\delta \ln(10)} \sqrt{\gamma_i^2 + \gamma_e^2 \frac{[1 - \delta(1-N)]^2}{2}} \quad (A24)$$

It can be shown that using the same approach for three replicates, i.e., considering $\varepsilon = (Z_1 + Z_2 + Z_3)/3$ and thus including the covariances $\sigma(Z_1, Z_2)$, $\sigma(Z_1, Z_3)$ and $\sigma(Z_2, Z_3)$, results in

$$\sigma_\varepsilon = \frac{1}{\delta \ln(10)} \sqrt{\gamma_i^2 + \gamma_e^2 \frac{[1 - \delta(1-N)]^2}{3}} \quad (A25)$$

Thus, the expression for $\sigma_\varepsilon$ for $U$ replicate sorption points becomes

$$\sigma_\varepsilon = \frac{1}{\delta \ln(10)} \sqrt{\gamma_i^2 + \gamma_e^2 \frac{[1 - \delta(1-N)]^2}{U}} \quad (A26)$$

Equation A26 is likely also valid for $U$ values higher than 3, but this was not further checked. Thus, this lengthy derivation results in the relatively simple Equation A26.

Next, we move to the calculation procedure of the effective variance as proposed by Bolster & Tellinghuisen (2010; their Equation 6) based on the procedure derived by Clutton-Brock (1967). Applying this procedure to the log-log transformed Freundlich equation (Equation 5) yields the following 'effective variance' $(\sigma_{\log X, eff})^2$ of $\log(X)$ in the case of $U$ replicate measurements of $c_e$:

$$\sigma_{\log X, eff}^2 = \left(\frac{d\log(X_d)}{d\log(c_i)}\right)^2 \sigma_{\log c_i}^2 + \left(\frac{d\log(X_d)}{d\log(c_e)} + \frac{d\log(X_F)}{d\log(c_e)}\right)^2 \frac{\sigma_{\log c_e}^2}{U} \quad (A27)$$

where $X_d$ is the $X$ function based on the decrease in the concentration in the liquid phase (Equation 2) and $X_F$ is the $X$ function of the Freundlich equation (Equation 1). The $d\log(X_d)/d\log(c_e)$ term of Equation A27 is called the 'direct' contribution from the error in $\log(c_e)$ to $(\sigma_{\log X})^2$, and the $d\log(X_F)/d\log(c_e)$ term is called the 'indirect' contribution from this error to $(\sigma_{\log X})^2$, as is illustrated by Figure 1.

It follows from Equation 5 that $d\log(X_F)/d\log(c_e) = N$, and from Equation A9 and Equation 3, it follows that $d\log(X_d)/d\log(c_e) = (\delta - 1)/\delta$. Using the same approach as in Equation A9, it can be shown that $d\log(X_d)/d\log(c_i) = 1/\delta$. Using again Equation A5, the following result is obtained from Equation A27:

$$\sigma^2_{\log X,eff} = \frac{1}{(\ln(10))^2}\left[\left(\frac{1}{\delta}\right)^2 \gamma_i^2 + \left(\left|\frac{\delta-1}{\delta}\right| + N\right)^2 \frac{\gamma_e^2}{U}\right] \tag{A28}$$

The absolute value of $(\delta - 1)/\delta$ is used because $\delta - 1$ is negative, which would lead to a negative contribution to the variance from $d\log(X_d)/d\log(c_e)$. This result is impossible (as illustrated by Figure 1). Some rearranging then yields:

$$\sigma^2_{\log X,eff} = \frac{1}{(\delta\ln(10))^2}\left[\gamma_i^2 + \gamma_e^2 \frac{[1-\delta(1-N)]^2}{U}\right] \tag{A29}$$

Thus, indeed the $(\sigma_{\log X,eff})^2$ of Equation A27 is identical to the $\sigma_\varepsilon^2$ of Equation A26. This confirms that the approach of calculating the variance of $\varepsilon$ is identical to the effective variance approach. Furthermore, this result confirms that Equation A26 is also valid for $U$ values above 3. Admittedly, the derivation based on Equation A27 proceeds somewhat more efficiently than that based on Equation A1 because no covariances are involved when using Equation A27. However, it is instructive to demonstrate that both approaches are identical.

To illustrate the correctness of Equation A26, Monte Carlo simulations were made following the procedures described in the main article, using the three sets of parameter values described in Table A1. The simulated $\chi^2$ distribution (calculated with Equation 7) was compared to the theoretically correct distribution. Cases 1 and 3 are typical for pesticide sorption, and Case 2 is typical for phosphate sorption. Figure A1 shows that the simulated $\chi^2$ distributions corresponded well for all three cases with the theoretical distributions for $L$-2 degrees of freedom ('2' because of the two Freundlich parameters).

**Table A1.** Parameters for the three cases used to illustrate the correctness of Equation A26 by Monte Carlo simulations.

|  | Case 1 | Case 2 | Case 3 |
|---|---|---|---|
| $R$ (kg/L) | 0.5 | 0.04 | 1 |
| $K_F$ (L/kg) | 0.5 | 200 | 1 |
| $N$ (-) | 0.9 | 0.3 | 0.7 |
| $\gamma_i$ (%) | 0.5 | 2.5 | 2.5 |
| $\gamma_e$ (%) | 1 | 5 | 5 |
| Number of replicates $U$ | 3 | 2 | 1 |
| Number of concentration levels $L$ | 5 | 6 | 9 |
| Degrees of freedom for $\chi^2$ (i.e., $L–2$) | 3 | 4 | 7 |
| Initial concentration levels (mg/L) | 0.1-0.32-1 -3.2-10 | 0.2-0.5-2- 5-10-20 | 0.1-0.2-0.32- 0.7-1-2-3.2-7-10 |

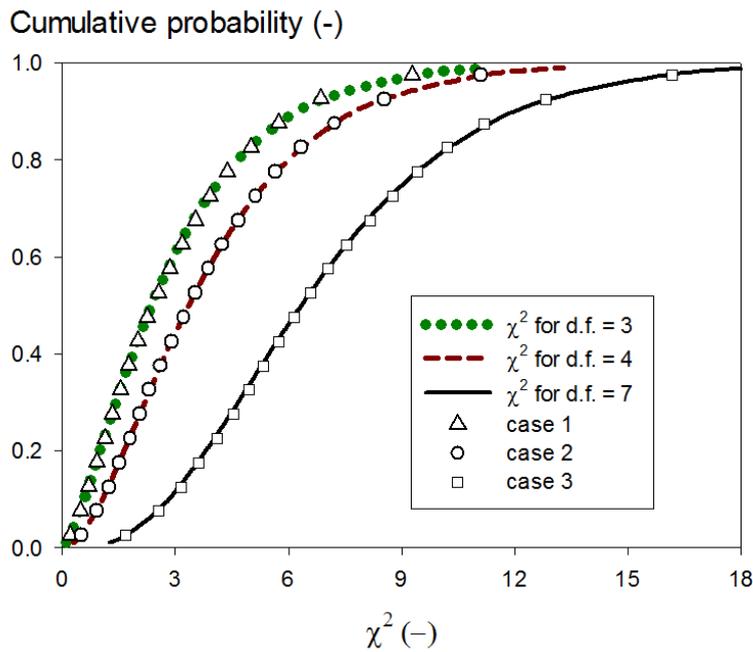

**Figure A1.** Comparison between the theoretical and simulated $\chi^2$ distributions of 10,000 fits to a sorption isotherm for the three cases with the parameters described in Table A1. The lines are the theoretical $\chi^2$ distributions for 3, 4 and 7 degrees of freedom, as indicated, and the symbols are percentiles of the simulated $\chi^2$ distributions.

# References

Anonymous. 2013. Covariance of transformed random variables; http://stats.stackexchange.com/questions/46874/covariance-of-transformed-random-variables. Accessed 02/03/2016.

Arfken, G.B. & Weber, H.J. 2005. Mathematical methods for physicists. Sixth edition, Elsevier Academic Press, Amsterdam.

Bolster, C.H. & Tellinghuisen, J. 2010. On the significance of properly weighting sorption data for least square analysis. Soil Sci. Soc. Am. J. 74, 670-679.

Clutton-Brock, M. 1967. Likelihood distributions for estimating functions when both variables are subject to error. Technometrics 9, 261–269.

Tellinghuisen, J. & Bolster, C.H. 2010. Least-squares analysis of phosphorus soil data with weighting from variance function estimation: a statistical case for the Freundlich isotherm. Environ. Sci. Technology 44, 5029-5034.